\documentclass[floatfix,aps,prd,showpacs,amsmath,nofootinbib,preprintnumbers,twocolumn]{revtex4}

\usepackage{graphicx}
\usepackage{bm}

\newcommand{\figurewidth}{3.4in}

\def\half{{1\over 2}}

\def\half{{1\over 2}}
\def\({\left(}
\def\){\right)}
\def\[{\left[}
\def\]{\right]}

\def\e{\begin{equation}}
\def\q{\end{equation}}
\def\m{\begin{eqnarray}}
\def\n{\end{eqnarray}}

\begin{document}

\title{An analytic calculation of the growth index for $f(R)$ dark energy model}

\author{Qing-Guo Huang}\email{huangqg@itp.ac.cn}
\affiliation{State Key Laboratory of Theoretical Physics, Institute of Theoretical Physics, 
Chinese Academy of Science, Beijing 100190, People's Republic of China}

\date{\today}

\begin{abstract}

We derive the analytic formula of the growth index for $f(R)$ dark energy model where the effect on the growth of matter density perturbation $\delta_m$ from modified gravity (MG) is encoded in the effective Newton coupling constant $G_{\rm eff}$ in MG (or equivalently $g\equiv {G_{\rm eff}/ G}$). Based on the analytic formula, we propose that the parameter $g$ can be directly figured out by comparing the observed growth rate $f_g\equiv d\ln\delta_m/d\ln a$ to the prediction of $f_g$ in general relativity.

\end{abstract}

\pacs{98.80.-k, 95.36.+x}

\maketitle


\section{Introduction}

The accelerating expansion of present Universe was discovered by type Ia supernovae \cite{Perlmutter:1998np,Riess:1998cb}. Up to now, the standard $\Lambda$CDM model in the framework of general relativity (GR) is able to explain the present cosmic acceleration within observational errors. However how to explain the tiny value of the cosmological constant compared to the known physical scales is still a big challenge.

Modified gravity (MG), for example the $f(R)$ gravity, provides a geometrical origin to the present cosmic acceleration. The basic idea of MG dark energy is that gravity is modified on the cosmological scales when the Ricci scalar $R$ is of order of today's Ricci scalar $R_0$, while GR is recovered in the region of $R\gg R_0$. However it is quite non-trivial to construct a viable $f(R)$ dark energy model which is consistent with both cosmological and local gravity constraints. See some typical viable $f(R)$ dark energy models in \cite{Amendola:2006we,Hu:2007nk,Appleby:2007vb,Starobinsky:2007hu,Tsujikawa:2007xu,Cognola:2007zu,Linder:2009jz}. 
It is useful to introduce the effective equation of state parameter $w=p_{\rm de}/\rho_{\rm de}$ to describe the difference between Friedmann-Robertson-Walker (FRW) background evolutions of MG and the standard $\Lambda$CDM model, where the effective pressure $p_{\rm de}$ and energy density $\rho_{\rm de}$ are determined by using the Einsteinian representation of gravitational field equations. On the other hand, since the gravity in MG is different from GR, the evolution of the matter density perturbation $\delta_m\equiv \delta\rho_m/\rho_m$ provides a crucial tool to distinguish MG dark energy model from dark energy model in GR, in particular the standard $\Lambda$CDM model. For simplicity, the growth rate $f_g$ of matter density perturbation can be parametrized by, \cite{Linder:2007hg}, 
\m
f_g\equiv {d\ln \delta_m\over d\ln a}\equiv \Omega_m(z)^{\gamma(z)},
\label{dgamma}
\n
where $a$ is the scale factor, $\Omega_m(z)$ is the density parameter for dust-like matter at redshift $z$, and $\gamma(z)$ is the so-called growth index. In $\Lambda$CDM model in GR, $w=-1$ and 
\m
\gamma\simeq 6/11, 
\n
\cite{Wang:1998gt,Linder:2005in}.

Generically the effect on the matter density perturbation in MG is encoded in the effective Newton coupling constant $G_{\rm eff}$. For simplicity, we introduce a new quantity $g\equiv G_{\rm eff}/G$ to measure the difference between MG and GR. In general, $w$ is time dependent and $g$ is time and scale dependent in MG, and then the growth index $\gamma$ is expected to be time and scale dependent. During deep matter dominant era GR is recovered, while the gravity is modified in the low redshift era when the cosmic acceleration occurs. One can expect that the evolutions of both FRW background and matter density perturbation in MG are too complicated to be solved analytically from the deep matter dominant era to accelerating era. 

In this paper we focus on the growth of matter density perturbation in the $f(R)$ dark energy model. We suppose that $g$ is parametrized as follows 
\m
g=g_0+g_1(1-\Omega_m), 
\label{paramg}
\n
where $g_0$ and $g_1$ are two constants. Here $g_1$ is used to characterize the time-evolution of $g$. Note that both $g_0=g_0(k)$ and $g_1=g_1(k)$ are scale dependent generically. In the deep matter dominant era $(\Omega_m\rightarrow 1)$, GR should be recovered and then $g\rightarrow 1$. But $g$ can deviate from one at low redshift. This parameterization can cover many viable $f(R)$ dark energy models at low redshift. Based on such a parameterization, we analytically solve the equation of motion of $\delta_m$ and work out an analytic formula of the growth index. 
Furthermore, we find that $g$ can be directly figured out by comparing the observed growth rate $f_g$ to the prediction of $f_g$ in GR.

This paper will be organized as follows. In Sec.~2 we briefly review the $f(R)$ dark energy model. In Sec.~3 we analytically calculate the growth index for $f(R)$ dark energy model. Summary and discussion are given in Sec.~4.

\section{A brief introduction to the $f(R)$ dark energy model}

Let's start with the following action 
\m
S={1\over 16\pi G}\int d^4 x \sqrt{-g} f(R)+S_m, 
\n
where $G$ is the Newton coupling constant, $S_m$ is the action for the matter, $R=6(2H^2+\dot H)$ and $H$ denotes the Hubble parameter. If $f(R)=R-2\Lambda$, the above action reduces to the Einstein-Hilbert action for the $\Lambda$CDM model in GR. In this paper we consider $f(R)$ vanishes for $R=0$, which implies that no cosmological constant is introduced. 
The $f(R)$ gravity contains a new scalar degree of freedom dubbed ``scalaron" whose mass depends on the Ricci scalar $R$ \cite{Starobinsky:1980te}. 
The stability of $f(R)$ theory requires 
\m
F\equiv f_{,R}>0,\quad F_{,R}\equiv f_{,RR}>0, 
\n
where $f_{,R}=d f(R)/dR$ and $F_{,R}=d F(R)/dR$. 
The former condition implies that gravity is attractive and graviton is not a ghost,  and the latter condition means that scalaron is not a tachyon. 
In addition, the viable $f(R)$ dark energy model is required to be similar to the $\Lambda$CDM model during the radiation and deep matter dominant era, but important observable deviations from the $\Lambda$CDM model appears at low redshift. In order to measure such a deviation, we can introduce a dimensionless quantity defined by $\beta\equiv {RF_{,R}/ F}$ 
which satisfies $0<\beta<1$ \cite{Amendola:2006we,Tsujikawa:2009ku}.

Considering $S_m$ describes the dust-like matter (the pressure of dust-like matter equals zero), the equations of motion for the FRW background take the form 
\m
H^2&=&{1\over 3}\[\half (FR-f)-3H\dot F-3(F-1)H^2\] \nonumber \\
&+&{8\pi G\over 3}\rho_m, \label{h2}\\
-2\dot H&=&\ddot F-H\dot F+2(F-1)\dot H+8\pi G \rho_m. \label{dh}
\n
Here we focus on the late time Universe where the radiation can be ignored. 
From these two equations, the effective energy density and pressure of $f(R)$ dark energy are respectively given by
\m
\rho_{\rm de}&=&{1\over 8\pi G}\[\half (FR-f)-3H\dot F-3(F-1)H^2\], \\
p_{\rm de}&=&-\rho_{\rm de}+{1\over 8\pi G}\[\ddot F-H\dot F+2(F-1)\dot H\], 
\n
and then the effective equation of state parameter $w$ reads 
\m
w=-1+{\ddot F-H\dot F+2(F-1)\dot H\over \half (FR-f)-3H\dot F-3(F-1)H^2}.
\n
Combining Eqs.~(\ref{h2}) and (\ref{dh}), the Ricci scalar becomes  
\m
R=3\[1-3w(1-\Omega_m)\] H^2, 
\n
where 
\m 
\Omega_m\equiv {8\pi G\rho_m\over 3H^2}
\label{domegam}
\n 
is the density parameter for the dust-like matter.

Many typical viable $f(R)$ dark energy models which are consistent with both cosmological and local gravity constraints are summarized in \cite{Tsujikawa:2009ku}. All of them can be written in the following form
\m
f(R)=R-\lambda R_s Y(x), 
\n
where $x=R/R_s$, $R_s (>0)$ is a characteristic value of $R$ and $\lambda$ is a positive parameter. The function $Y(x)$ in the viable model takes the form: (i) $Y(x)=x^p\ (0<p<1)$ \cite{Amendola:2006we}, (ii) $Y(x)=x^{2n}/(x^{2n}+1)\ (n>0)$ \cite{Hu:2007nk}, (iii) $Y(x)=1-(1+x^2)^{-n}\ (n>0)$ \cite{Starobinsky:2007hu}, (iv) $Y(x)=1-e^{-x}$ \cite{Cognola:2007zu,Linder:2009jz}, (v) $Y(x)=\tanh(x)$ \cite{Tsujikawa:2007xu}, etc. We find that all of these models satisfy $F=f_{,R}<1$.

\section{The analytic formula of the growth index for $f(R)$ dark energy model}

From now on, we will focus on the evolution of matter density perturbation in MG. 
In the sub-horizon limit, the evolution of matter density fluctuation $\delta_m$ is govern by,   
\m
\ddot \delta_m+2H\dot \delta_m-4\pi G_{\rm eff}\rho_m \delta_m=0,
\label{delta}
\n
\m
G_{\rm eff}=g(a,k,R)\cdot G, 
\n
where 
\m
g(a,k,R)\equiv {1\over F}\(1+{1\over 3}{1\over 1+{M^2a^2\over k^2}}\), 
\label{defg}
\n
in \cite{Zhang:2005vt,Tsujikawa:2007gd}, 
or without taking any approximation at the matter-dominant stage \cite{delaCruzDombriz:2008cp,Motohashi:2009qn}
\m
g(a,k,R)=1+{8\over 3}{4/F-3\over 27+8\({k^2\over a^2M^2}\)^4} \({k^2\over a^2M^2}\)^4, 
\label{defg2}
\n
and
\m
M^2={R\over 3\beta}={F\over 3F_{,R}}.
\n
Here $M$ is nothing but the mass of scalaron. The positivities of both $f_{,R}$ and $f_{,RR}$ guarantee the positivity of the mass square of scalaron. In the scales which are much smaller than $M^{-1}$, GR is recovered, and the gravity is modified in the scales around or larger than $M^{-1}$. In addition, considering (\ref{defg}) or (\ref{defg2}) with $F<1$, we find $g>1$ for the viable $f(R)$ dark energy models in the literatures.


In the deep matter dominant era, $a\sim t^{2/3}$ and then Eq.~(\ref{delta}) becomes 
\m
\ddot \delta_m+{4\over 3t}\dot \delta_m-{2\over 3t^2} g \delta_m=0,
\n
whose solution is given by $\delta_m\sim t^{\sqrt{1+24g}-1\over 6}\sim a^{\sqrt{1+24g}-1\over 4}$, 
where $g$ is taken as a constant. 
In this era, GR is proposed to be recovered ($g\rightarrow 1$) and then $\delta_m\sim a$. 


Now let's switch to the late time Universe where the energy densities of effective dark energy and dust-like matter are comparable to each other. Eq.~(\ref{delta}) can be re-written as follows  
\m
&&{d^2 \ln \delta_m\over d\ln a^2}+\({d\ln \delta_m\over d\ln a}\)^2+{d\ln \delta_m\over d\ln a}\[\half-{3\over 2}w(1-\Omega_m)\] \nonumber \\
&&={3\over 2}g\Omega_m, 
\label{ddelta}
\n
or equivalently 
\m
{df_g\over d\ln a}+f_g^2+f_g\[\half-{3\over 2}w(1-\Omega_m)\]={3\over 2}g\Omega_m. 
\label{dfg}
\n
From Eqs.~(\ref{h2}), (\ref{dh}) and (\ref{domegam}), we have 
\m
{d\Omega_m\over d\ln a}=3w \Omega_m (1-\Omega_m). 
\n
Combining with the definition of the growth index $\gamma$ in Eq.~(\ref{dgamma}), Eq.~(\ref{ddelta}) becomes 
\m
&&3w\Omega_m (1-\Omega_m)\ln \Omega_m {d\gamma\over d\Omega_m}+3w(\gamma-\half)(1-\Omega_m) \nonumber \\
&&+\Omega_m^\gamma-{3\over 2}g\Omega_m^{1-\gamma}+\half=0. 
\label{dog}
\n

Usually the form of $g(a,k,R)$ is expected to be very complicated. In order to capture the main feature of $f(R)$ dark energy model at low redshift, we expand $g$ as a power series about $(1-\Omega_m)\sim 0$ for a given perturbation mode $k$, 
\m
g=\sum_{n=0} g_n (1-\Omega_m)^n. 
\n
In this paper we take the first two terms like that in Eq.~(\ref{paramg}) into account. \footnote{The case with $g_0=1$ is discussed in \cite{Ferreira:2010sz}. }
For slowly varying equation of state parameter $w$ $\(|dw/d\Omega_m|\ll (1-\Omega_m)\)$, the solution of Eq.~(\ref{dog}) takes the form  
\m
\gamma={c_{-1}\over 1-\Omega_m}+c_0+c_1 (1-\Omega_m)+{\cal O}\((1-\Omega_m)^2\), 
\label{afgamma}
\n
where $c_{-1}$, $c_0$ and $c_1$ can be calculated order by order, 
\m
c_{-1}&=&\ln {1+\sqrt{1+24g_0}\over 6g_0}, \\
c_0&=&1-{c_{-1}\over 2}+{-2+3e^{c_{-1}}w-3e^{2c_{-1}}g_1\over 2+3e^{2c_{-1}}g_0-6e^{c_{-1}}w}, \\
c_1&=&{(c_{-1}^2+4c_{-1}c_0-4(1-c_0)c_0)(2-3e^{2c_{-1}}g_0) \over 8(2+3e^{2c_{-1}}g_0-12e^{c_{-1}}w)} \nonumber \\
&-&{c_{-1}(4-3e^{2c_{-1}}g_0-6e^{c_{-1}}w) \over 6(2+3e^{2c_{-1}}g_0-12e^{c_{-1}}w)}\nonumber \\
&+&{3g_1(1-c_{-1}/2-c_0)e^{2c_{-1}}\over 2+3e^{2c_{-1}}g_0-12e^{c_{-1}}w}.
\n 
The expressions of the higher order terms are quite complicated, and the readers can easily work them out once they need. 
Here the first two terms on the right hand side of Eq.~(\ref{afgamma}) make main contributions to $\gamma$ and the term with $c_1$ is roughly negligible if both $(g_0-1)$ and $g_1$ are much less than one. 
If $g_0\neq 1$, the growth index is expected to be time-evolving, and the ansatz with a constant growth index is not generic for $f(R)$ dark energy model. Our analytic formula indicates that a better ansatz for $\gamma$ is 
\m
\gamma(z)\simeq {\gamma_{-1}\over 1-\Omega_m}+\gamma_0+\gamma_1(1-\Omega_m),
\n
where $\gamma_{-1}$, $\gamma_0$ and $\gamma_1$ are constants.

For $g_0=1$ and $g_1=0$, our result reduces to GR where $c_{-1}=0$, $c_0={3(1-w)\over 5-6w}$ and $c_1={3\over 125}{(1-w)(1-3w/2)\over (1-6w/5)^2(1-12w/5)}$. \footnote{The denominator of $c_1$ is slightly different from that in \cite{Wang:1998gt} where $c_1={3\over 125}{(1-w)(1-3w/2)\over (1-6w/5)^3}$. } 
For $g_0=1$, 
\m
c_{-1}=0,\quad c_0={3(1-g_1-w)\over 5-6w}.
\n 
In Dvali-Gabadadze-Porrati (DGP) model \cite{Dvali:2000hr,Deffayet:2001pu}, $g=1-{1\over 3}{1-\Omega_m^2\over 1+\Omega_m^2}$. In the matter dominant era, $g\rightarrow 1-{1\over 3}(1-\Omega_m)$ and $w\rightarrow -1/2$, and thus $\gamma\simeq {11/16}$ which is the same as that in \cite{Linder:2007hg}.

Nowadays the property of dark energy has been tightly constrained from observations \cite{Ade:2013zuv}. $\Lambda$CDM model can fit the data, and the room for $f(R)$ dark energy model has been tightly constrained, for example $|g_0-1|\lesssim {\cal O}(0.1)$ and $|g_1|\lesssim {\cal O}(0.1)$. Therefore $c_{-1}$ and $c_0$ can be expanded around the case of GR ($g_0=1$ and $g_1=0$), 
\m
c_{-1}&=&-{3\over 5}(g_0-1)+{\cal O}((g_0-1)^2), \\
c_0&=&{3(1-g_1-w)\over 5(1-6w/5)} \\
&+&{9(7+18g_1-20w-12g_1 w+12w^2)\over 250(1-6w/5)^2}(g_0-1)  \nonumber \\
&+&{\cal O}((g_0-1)^2). \nonumber
\n
For $w=-1$ and $g_1=0$, $c_0\simeq {6\over 11}+{351\over 1210}(g_0-1)$.

Applying our analytic formula in (\ref{afgamma}) to Eq.~(\ref{dgamma}), we can easily calculate the growth rate $f_g(z)$. 
Testing growth rate $f_g$ from our analytic result in (\ref{afgamma}) against the value obtained by numerical calculation, the accuracy at low redshift $(z\lesssim 1)$ is better than a few percents. 
Combining Eq.~(\ref{afgamma}) with Eq.~(\ref{dgamma}) and expanding $f_g$ up to the order of $(1-\Omega_m)^2$, we have 
\m
f_g(z)=e^{-c_{-1}-({c_{-1}/ 2}+c_0)(1-\Omega_m)+{\cal O}\((1-\Omega_m)^2\)}, 
\n
where 
\m
{c_{-1}\over 2}+c_0&=& {3(1-g_1-w)\over 5(1-6w/5)} \nonumber \\
&-&{3(2-27g_1+18g_1 w)\over 125(1-6w/5)^2}(g_0-1) \nonumber \\
&+&{\cal O}((g_0-1)^2). 
\label{cc}
\n
Since the first term on the right hand side of Eq.~(\ref{cc}) is dominant, we have 
\m
{f_{g,\rm{MG}}(z)\over f_{g,{\rm GR}}(z)}\simeq \exp\[-c_{-1}+{3g_1\Omega_{\rm de}(z)\over 5-6w}\], 
\label{ddeltaa}
\n
where $\Omega_{\rm de}=(1-\Omega_m)$ is the dark energy density parameter. 
For $g_1=0$ and $g_0>1$ (or equivalently $c_{-1}<0$) which implies that gravity is stronger than GR, the growth rate of matter density perturbation is enhanced by a factor $e^{-c_{-1}}$.
Motivated by Eq.~(\ref{ddeltaa}), we propose  
\m
1-{5\over 3}c_{-1}+{g_1\Omega_{\rm de}(z)\over 1-6w/5}\simeq r_{g,{\rm obs}}(z), 
\label{rgz}
\n
where 
\m
r_{g,{\rm obs}}(z)\equiv 1+{5\over 3}\ln{f_{g,{\rm obs}}(z)\over f_{g,{\rm GR}}(z)}. 
\n
Note that $c_{-1}$ is a function of $g_0$.
Once we can construct the relation between $r_g(z)$ and $\Omega_{\rm de}$ from cosmological observations, we can easily figure out $g_0$ and $g_1$. Roughly speaking, the value of $g_0$ can be determined by the value of $r_{g,{\rm obs}}(z)$ when $\Omega_{\rm de}\simeq 0$, and $g_1$ is related to the tilt of $r_{g,{\rm obs}}(z)$ at low redshift. If $|g_0-1|\lesssim {\cal O}(0.1)$,  Eq.~(\ref{rgz}) becomes 
\m
g_0+{g_1\Omega_{\rm de}(z)\over 1-6w/5}\simeq r_{g,{\rm obs}}(z), 
\n
which indicates that the redshift-independent part of $r_{g,{\rm obs}}(z)$ is equal to $g_0$.

\begin{figure}[hts]
\centerline{\includegraphics[width=\figurewidth]{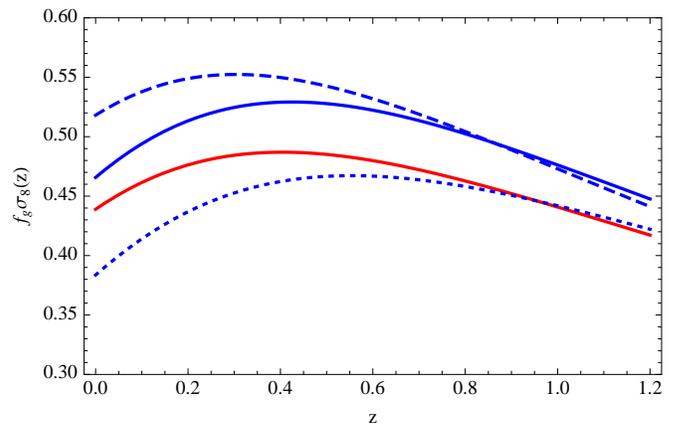}}
\caption{The plot of $f_g\sigma_8(z)$. Here we adopt $w=-1$, $\Omega_m^0=0.315$ and $\sigma_8=0.829$. The red solid, blue dashed, blue dotted and blue solid curves correspond to $\Lambda$CDM model, $g=1.3$, $g=1-0.5(1-\Omega_m)$ and $g=1.3-0.5(1-\Omega_m)$ respectively. }
\label{fig:fs8}
\end{figure}

In the literatures, one may prefer to constrain $f_g\sigma_8(z)$ from cosmological observations, where $\sigma_8$ is today's root-mean-square mass fluctuation on $8h^{-1}$ Mpc. Because the equation of motion of $\delta_m$ is a linear equation, one can define a normalized growth function $D(z)$ via 
\m
D(z)\equiv \delta_m(z)/\delta_m(z=0), 
\n
and then $\sigma_8(z)=\sigma_8 D(z)$. Solving Eq.~(\ref{dgamma}), we get 
\m
D(z)={1\over 1+z}\exp \[\int_0^z \(1-\Omega_m(z')^{\gamma(z')}\){dz'\over 1+z'}\].
\n
Therefore 
\m
f_g\sigma_8 (z)=D(z)\Omega_m(z)^{\gamma(z)}\sigma_8. 
\n
Using our analytic result, $f_g\sigma_8(z)$ is plotted in Fig.~\ref{fig:fs8}. 
Roughly speaking, if $|g_0-1|\lesssim 1$ and $|g_1|\lesssim 1$, $g_0$ shifts the amplitude of $f_g\sigma_8(z)$ and $g_1$ changes the shape of $f_g\sigma_8(z)$.


\section{Summary and discussion}

To summarize, we analytically calculate the growth index in the $f(R)$ dark energy model. Actually our results are applicable for more general MG dark energy models, for example $f(T)$ dark energy model \cite{Linder:2010py,Bamba:2010ws,Zheng:2010am,Zhang:2012jsa}, as long as the effect on the growth of matter density perturbation from MG is encoded in $g=G_{\rm eff}/G$.  
As we know, there are two key parameters for MG dark energy model, namely $w$ and $G_{\rm eff}$ (or equivalently $g$). The former parameter determines the expansion history of our Universe, and the latter parameter tells us how the matter density perturbation grows up. 
Adopting the analytic formula, we find a simple relation between $g$ and the growth rate in Eq.~(\ref{ddeltaa}), and then we propose that $g$ can be directly figured out by comparing the observed growth rate $f_g$ to the prediction of $f_g$ in GR. In the literatures, one would like to use $f_g\sigma_8(z)$ to characterize the growth of matter density perturbation. In this case one can also use our analytic formula to calculate $f_g\sigma_8(z)$ and then fit out $g_0$ and $A$ from the data.

Recently the anisotropic clustering of the Baryon Oscillation Spectroscopic Survey (BOSS) CMASS Data Release 11 (DR11) sample was analyzed. 
The combination of Planck and CMASS implies $\gamma=0.772_{-0.097}^{+0.124}$ and a similar result $\gamma=0.76\pm 0.11$ is obtained when replacing Planck with WMAP9 in \cite{oai:arXiv.org:1312.4611}. Both results deviate from the prediction of $\Lambda$CDM in GR at more than $2\sigma$ level. 
The large value of $\gamma$ may come from the the large value of $\sigma_8$ from Planck, or it is just a statistical fluctuation. 
Considering $f_g\sigma_8(z=0.57)=0.419\pm 0.044$ from BOSS CMASS DR11, we obtain $g_0\simeq 0.73$ in the reference $\Lambda$CDM model ($\Omega_m^0=0.315$ and $\sigma_8=0.829$) from Planck \cite{Ade:2013zuv}. A careful data fitting will be done in the near future \cite{huang2014}. 
In a word, if such a deviation is confirmed in the future, we really need to modify the gravity. 

Finally see some other aspects on $f(R)$ dark energy model in \cite{DeFelice:2010aj,Nojiri:2010wj,Polarski:2007rr,Wei:2008ig,Gong:2008fh,Gannouji:2008wt,Bamba:2008hq,Wu:2009zy,Biswas:2011ar,Bamba:2012qi,He:2012rf,Basilakos:2013nfa,Abebe:2013zua,Xu:2013tsa,Dossett:2014oia,Pouri:2014nta} etc.

\noindent {\bf Acknowledgments}

This work was initiated during High1-2014 KIAS-NCTS joint workshop on particle physics, string theory and cosmology. Q.-G.H. is supported by the project of Knowledge Innovation Program of Chinese Academy of Science and grants from NSFC (grant NO. 10821504, 11322545 and 11335012).



\end{document}